\documentclass[12pt ]{article}
\usepackage{amsmath}
\usepackage{graphicx}
\usepackage{hyperref}
\usepackage{lmodern}
\usepackage{amssymb}
\usepackage{booktabs}
\usepackage{tikz}
\usetikzlibrary{patterns}
\usepackage{pdfpages}

\renewcommand{\theequation}
{\arabic{section}.\arabic{equation}}

\makeatletter
\def\eqnarray{ \stepcounter{equation} \let\@currentlabel=\theequation
 \global\@eqnswtrue
 \global\@eqcnt\z@
 \tabskip\@centering
 \let\\=\@eqncr
 $$\halign to \displaywidth\bgroup\@eqnsel\hskip\@centering
 $\displaystyle\tabskip\z@{##}$&\global\@eqcnt\@ne
 \hfil$\displaystyle{{}##{}}$\hfil
 &\global\@eqcnt\tw@$\displaystyle\tabskip\z@{##}$\hfil
 \tabskip\@centering&\llap{##}\tabskip\z@\cr}
\makeatother

\newcommand{\be}{\begin{equation}}
\newcommand{\ee}{\end{equation}}

\newcommand{\beqa}{\begin{eqnarray}}
\newcommand{\eeqa}{\end{eqnarray}}
\newcommand{\nn}{\nonumber}

\def\CG {{\cal G}}

\begin{document}

\setlength{\baselineskip}{7mm}
\begin{titlepage}
 
\begin{flushright} 
 {\tt NRCPS-HE-72-2024} 
\end{flushright}

\begin{center}
{\Large ~\\{\it     Condensation of Magnetic Fluxes \\and\\ Landscape of QCD Vacuum\footnote{The presentations given at the Tor Vergata University on October 9, 2014, Rome, Italy and at the International Conference on "QCD Vacuum Structure and Confinement"  (QCD-VSC-2024), Naxos, Greece.} 

}

}

\vspace{3cm}

{\sl George Savvidy

\centerline{${}$ \sl Institute of Nuclear and Particle Physics}
\centerline{${}$ \sl Demokritos National Research Center, Ag. Paraskevi,  Athens, Greece}

}
 
\end{center}
\vspace{3cm}

\centerline{{\bf Abstract}}

I discuss new non-perturbative solutions of the sourceless Yang-Mills equation representing the superposition of oppositely oriented chromomagnetic flux tubes (vortices) similar in their form to a lattice of superposed Abrikosov-Nielsen-Olesen chromomagnetic vortices.  These solutions represent highly degenerate classical vacua of the Yang Mills theory that are separated by potential barriers and are forming a complicated potential landscape of the QCD vacuum. It is suggested that the solutions describe a lattice of dense chromomagnetic  vortices representing a dual analog of the Cooper pairs condensate in a superconductor.

  \vspace{12pt}

\noindent

\end{titlepage}

\pagestyle{plain}

\section{\it Introduction }
The pure Yang Mills theory does not have adjoint scalar fields and therefore does not admit explicit 't Hooft-Polyakov monopole solutions or Nielsen-Olesen magnetic flus tubes.  However, it was largely expected that the confining QCD vacuum should represent a dual superconductor state with a condensate of the monopole-antimonopole pairs in analogy with Cooper pairs in ordinary superconductors  \cite{  tHooft:1981bkw, tHooft:1979rtg, Mandelstam:1978ed, tHooft1977,  Mandelstam:1980ii,cmp/1104178138, https://doi.org/10.1112/plms/s3-55.1.59, Gukov:2006jk, Gukov:2008sn, Kapustin:2005py, Kapustin:2006pk, DeBenedictis:2024dts}.   Whether or not there is a  condensation of the monopole-antimonopole pairs or of the magnetic flux tubes depends on the details of the dynamics, on the  existence of solutions with long-range topological structure in the pure Yang Mills theory.  The dynamics of magnetic flux tubes must be an important ingredient of the confinement mechanism in addition to the  topological arguments.  The question is if such long-range topological structure exists in the pure Yang Mille theory,  in QCD.

Here we investigate  a new class of exact vacuum solutions of the sourceless Yang Mills equation, which have nontrivial topological structure  and singularities that are distributed over two-dimensional sheets and cylinders representing a lattice of superposed vortices of magnetic flux tubes \cite{Savvidy:2024sv, Savvidy:2024ppd, Savvidy:2024xbe}.   These singularities of the gauge potential do not show up in the field strength tensor, which  has a regular behaviour.  The singularities of the gauge potential are located on the planes that are invariant with respect to the nonsingular gauge transformations and characterise the moduli space of solutions. 
 
 There is an appealing analogy with the Schwarzschild solution in gravity, where the solution is asymptotically flat and regular at infinity but has metric singularity at the horizon $ r_g=2MG/c^2$. This singularity is not a real one because the Riemann curvature tensor is regular at   $r_g$. In our case as well we have singularities of the gauge field while the field strength tensor - the curvature of the Yang Mills theory - is regular.  The moduli space of vacuum solutions is characterised by 3d-vectors, and we demonstrated that these vacuum configurations are separated by potential barriers forming a complicated landscape of the QCD vacuum.

\section{\it Sourceless gauge fields }

The well known solution of the covariantly constant  equation 
\be\label{YMeqcov}
\nabla^{ab}_{\rho} G^{b}_{\mu\nu} =0
\ee 
has the following form \cite{Batalin:1976uv, Savvidy:1977as, Matinyan:1976mp, Brown:1975bc, Duff:1975ue}: 
 \be\label{consfield}
A^{a}_{\mu} = - {1\over 2} F_{\mu\nu} x_{\nu}   n^a , 
\ee
where $F_{\mu\nu} $ and $n^a$ are space-time independent parameters, $ n^{a} n^{a} =1$ and 
$
G^{a}_{\mu\nu} =  \partial_{\mu} A^{a}_{\nu} - \partial_{\nu} A^{a}_{\mu}
 - g \varepsilon^{abc} A^{b}_{\mu} A^{c}_{\nu},
$ 
$
\nabla^{ab}_{\mu}(A)=
  \delta^{ab} \partial_{\mu}   - g \varepsilon^{acb} A^{c}_{\mu}.
$
They are the solutions of the sourceless  Yang-Mills equation $\nabla^{ab}_{\mu} G^{b}_{\mu\nu} =0$ as well. 
The new solutions can be obtained through the nontrivial space-time dependence of the unit vector $n^a(x)$ \cite{Savvidy:2024sv, Savvidy:2024ppd, Savvidy:2024xbe}.  Considering the Ansatz  \cite{Cho:1979nv, Cho:1980nx, Cho:2010zzb, tHooft:1974kcl, Corrigan:1975zxj, Biran:1987ae}
\be\label{choansatz}
A^{a}_{\mu} =  B_{\mu} n^{a}  +
{1\over g} \varepsilon^{abc} n^{b} \partial_{\mu}n^{c},
\ee
where $B_{\mu}(x)$  is the Abelian Lorentz vector and $n^{a}(x)$ is a space-time dependent colour unit vector
$
n^{a} n^{a} =1,
$
$n^{a} \partial_{\mu} n^{a} =0,
$
one can observe that the field strength tensor factorises  \cite{Cho:1979nv, Cho:1980nx}:
\be\label{chofact}
G^{a}_{\mu\nu} =  ( F_{\mu\nu} + {1\over g} S_{\mu\nu})~ n^{a} \equiv G_{\mu\nu}~ n^a,
\ee
where
\be\label{spacetimefields}
F_{\mu\nu}= \partial_{\mu} B_{\nu} - \partial_{\nu} B_{\mu},~~~~~~~~~
S_{\mu\nu}= \varepsilon^{abc} n^{a} \partial_{\mu} n^{b} \partial_{\nu} n^{c}.
\ee
The general solution of the equation (\ref{YMeqcov}) depends on an arbitrary function $\theta(X) $  of  the variable $X= a_1 x +a_2 y + a_3 z-a_0 t$, thus
$
\theta(X)=  \theta(a_1 x +a_2 y + a_3 z-a_0 t )=\theta( a\cdot x ),
$
where $a_{\mu}, \mu=1,2,3,0$ are arbitrary real numbers and of the second variable $Y=b_1 x +b_2 y + b_3 z-b_0 t= ( b\cdot x )$. We have the following solution for the colour unit vector \cite{Savvidy:2024sv, Savvidy:2024ppd, Savvidy:2024xbe}:
\be\label{generasol}
n^a(\vec{x})= \{\sin \theta(X)  \cos\Big({Y \over \theta(X)^{'} \sin\theta(X)}\Big),~\sin\theta(X) \sin\Big({Y \over \theta(X)^{'}  \sin\theta(X)}\Big),~ \cos\theta(X)   \}.
\ee 
 The explicit form of the vector potential $A^a_{\mu}$ can be obtained by substituting the  unit colour vector (\ref{generasol}) into (\ref{choansatz}), and the result is given in (\ref{magneticsheetsolution1}).  The arbitrary function $\theta(X)$ in the equation (\ref{generasol}) defines the moduli space of the solutions.  The singularities are located on the planes $X_s$ where the  $\sin\theta(X)$  vanishes:  \be\label{singplanes} 
\theta(X_s) = 2 \pi N, ~~~N=0,\pm 1, \pm 2...., ~~~~~~\theta(X_s)^{'} \neq 0
\ee
and  $\cos/\sin\Big({Y \over \theta(X)^{'} \sin\theta(X)}\Big)$ are fast oscillating trigonometric functions. Despite the fact that the first term $\partial_{\mu} G^a_{\mu\nu}$ in the Yang Mills equation is in general singular it appears that the second term $g \epsilon^{abc} A^b_{\nu} G^c_{\nu\mu}$ also has singularities that compensate each other, and the equation is fulfilled in the vicinity of the singular planes. Their positions are invariant with respect to the nonsingular gauge transformations and therefore intrinsically characterise the moduli space of the solution.

There is an appealing analogy with the Schwarzschild solution in gravity, where the solution is asymptotically flat and regular at infinity while it  has metric singularity at the Schwarzschild radius $r \rightarrow r_g=2MG/c^2$, at the radius that defines the event horizon. This metric singularity is not a physical one because the Riemann curvature tensor is regular at $r=r_g$, and there are rich physical phenomena that are developing at the Schwarzschild radius. 

In our case we have singularities of the gauge field, while the field strength tensor - the curvature of the Yang Mills theory - is regular and one can expect new physical phenomena to appear at these locations, as we shall demonstrate below. These non-perturbative solutions of the sourceless Yang-Mills equation represent a lattice of superposed chromomagnetic vortices of oppositely oriented magnetic fluxes, and they were described in \cite{Savvidy:2024sv, Savvidy:2024ppd, Savvidy:2024xbe}.

There are two subclasses of physically interesting solutions: the time independent solutions $a_0 = 0$ describing stationary magnetic fluxes distributed in 3d-space and time dependent solutions $a_0 \neq 0$ describing propagation of strings and branes. The general solution (\ref{generasol}) for the vector potential $A^a_{\mu}$   (\ref{choansatz})  depends on two coordinates $X$ and $Y$.  For the sake of transparency and compactness of the subsequent formulas we will identify this plane as the plane  $(x,y)$. Thus we are considering the vectors  $a_{\mu}=(0,a,0,0)$ and $b_{\nu}=(0,0,b,0)$, so that  $\theta(x) = f(a x)$, $\phi(x,y) = b y /f^{'}(ax) \sin f(ax)   $.  The gauge field (\ref{choansatz}) at  $B_{1}=H y$ will take the following form:
\beqa\label{magneticsheetsolution1}
A^{a}_{\mu}(x,y) &=&   {1\over g} \left\{
\begin{array}{ccccc}   
(0,0,0) \\
a \Big( b y ( ({g H \over a b}-1)  \sin f    +{1 \over \sin f  } )     \cos ({b y  \over f^{'}  \sin f }) - f^{'}  \sin ({b y  \over f^{'} \sin f }) +  b y  {  f^{''}   \over f^{'2}} \cos f  \cos ({b y  \over f^{'}  \sin f }), \\ 
~~ b y ( ({g H \over a b}-1)  \sin f    +{1 \over \sin f  } )      \sin({b y  \over f^{'}  \sin f }) + f^{'}   \cos({b y  \over f^{'}  \sin f }) + b y {  f^{''}   \over f^{' 2}} \cos f  \sin ({b y  \over f^{'}  \sin f}), ~~~\\
    b  y ({g H \over a b}-1) \cos f -   { f^{''}  \over f^{'2}}  \sin f  )\Big)  \\ 
{b \over f^{'}} \Big(-\cos f  \cos ({b y  \over f^{'} \sin f}),-\cos f  \sin ({b y  \over f^{'} \sin f}), ~  \sin f \Big)\\
(0,0,0)
\end{array} \right. ,  
\eeqa 
where the derivatives are over the whole argument $ax$ and singularities are at (\ref{singplanes}). One can verify explicitly  that it is a solution of the Yang Mills equation \cite{Savvidy:2024sv, Savvidy:2024ppd, Savvidy:2024xbe}. The nonzero component of the  field strength tensor $G^a_{\mu\nu}$ is of the  following form:
 \be\label{consfielstr}
 G^{a}_{12}(x) =  {a b - g H \over g}   ~ n^a(x,y),
 \ee
 and the energy density of the chromomagnetic field is without singularities and is a space-time constant:
 \be\label{energydenscont}
\epsilon =  {1\over 4 }G^{a}_{ij} G^{a}_{ij} =    { (gH-a b )^2 \over 2 g^2}.
 \ee  
There are two important limiting solutions when $H=0$  and $g H = a b$. In the first case we have the solution that has a constant energy density $\epsilon =   (a b )^2 / 2 g^2$,  like the solution (\ref{consfield}), and, as we shall demonstrate in the next section, are separated from each other by a potential barrier. In the second case we have new vacuum solutions with $G_{\mu\nu} =0$ and $\epsilon =0$ which are characterised by the vectors $(\vec{H}, \vec{a}, \vec{b})$ and are separated from the $A_{\mu}=0$ vacuum  by the potential barriers. In the next sections we shall analyse the shape of these barriers.

\section{\it Potential barriers between solutions of constant energy density }

 In order to investigate this potential landscape in the first case of constant energy density $H=0$, $\epsilon =   (a b )^2 / 2 g^2$ we shall  perform a gauge transformation $U(\vec{x})$ that transforms the unit colour vector $n^a(x)$ (\ref{generasol}) into the constant vector in the third direction $n^{'a} = (0,0,1)$:
\be
\hat{n}^{'} = U^- \hat{n} U,~~~~~~~\hat{n} = n^a  \sigma^a.
\ee
The $SU(2)$  matrix of the corresponding {\it singular gauge transformation} has the following form:
\beqa\label{singtruns}
U=  
\left(
\begin{array}{cc}
\cos({f \over 2}) e^{{i\over 2} ({\pi \over 2} - {b y \over  f^{'}\sin f})}&i \sin({f \over 2}) e^{{i\over 2} ({\pi \over 2} - {b y \over f^{'}\sin f})}\\
i \sin({f \over 2}) e^{-{i\over 2} ({\pi \over 2} - {b y \over f^{'} \sin f})}&\cos({f \over 2}) e^{-{i\over 2} ({\pi \over 2} - {b y \over f^{'} \sin f})}
\end{array}
\right)\nn,
\eeqa
which   transforms  the  gauge potential (\ref{magneticsheetsolution1}) into the constant  chromomagnetic field $A^{'}_{\mu}= U^- A_{\mu} U - {i\over g} U^- \partial_{\mu} U$  of the form   (\ref{consfield})  where only $A^{'3}_{1}$ is  nonzero
\be\label{magneticsheetsolution3}
A^{'3}_{1}=  {g H -a b\over g} y, ~~~~~G^{' 3}_{12} ={ ab - g H\over g},~~~~~~~\epsilon =   { (g H -a b)^2 \over 2 g^2}
\ee  
and we have to take $H=0$. Let us  consider an arbitrary path $w(\alpha)$ that joins the field configurations (\ref{magneticsheetsolution1}) and (\ref{magneticsheetsolution3})  and calculate the corresponding magnetic energy density. By multiplying  the potential (\ref{magneticsheetsolution1}) by the  factor $w({1\over 2}- \alpha)$ and the potential (\ref{magneticsheetsolution3}) by the factor $ w({1\over 2}+ \alpha)$  and requiring  that $w(0)=0$ and $w(1)=1$, when $\alpha \in [-{1\over 2}, +{1\over 2}]$ \cite{Jackiw:1976pf, Callan:1976je} :
\beqa\label{pathdeformation}\
\begin{array}{cccc}   
&\hat{A}^{a}_{0}= &w({1\over 2}- \alpha) A^{a}_{0}&\\
&\hat{A}^{a}_{1}= &w({1\over 2}- \alpha) A^{a}_{1}& +w({1\over 2}+ \alpha) A^{'a}_{1} \\
&\hat{A}^{a}_{2}= &w({1\over 2}- \alpha) A^{a}_{2}&   \\
&\hat{A}^{a}_{3}= &w({1\over 2}- \alpha) A^{a}_{3} &+w({1\over 2} + \alpha) A^{'a}_{3}.
\end{array}  
\eeqa By substituting the field $(\ref{pathdeformation})$ into the energy density functional $\epsilon =  {1\over 4 }G^{a}_{ij} G^{a}_{ij}$ we will get the following shape of the potential barrier:
\beqa
\epsilon(x,\alpha)&=& {a^2 b^2 \over 2 g^2} \Big( (2 -w_{-})^2 w^2_{-} + w^2_{+} +  
 2 (2 -w_{-})w_{-} (1 +w_{-} )w_{+}  \cos f(ax)+ {w^2_{-} w^2_{+} \over \sin^2 f(ax)}~ \Big),~~~~~~~
\eeqa
where $w_- \equiv  w({1\over 2}- \alpha)$ and   $w_+ \equiv w({1\over 2}+ \alpha)$.   If $w(\alpha)$ is a linear functional of its argument $w = {1\over 2} -\alpha$, then we will get 
\be\label{barriershape}
\epsilon(x,\alpha)= {a^2 b^2 \over 32 g^2} \Big( 12 - 8 \alpha + 16 \alpha^2 + 32 \alpha^3  +  (18 - 80 \alpha^2 + 32 \alpha^4) \cos f(ax) + {(1-4 \alpha^2)^2 \over \sin^2 f(ax)}~ \Big). 
\ee
At $\alpha = \pm 1/2$  the energy densities are equal to  $\epsilon = {a^2 b^2 \over 2 g^2} $ and there is a potential barrier between these field configurations $A_{\mu}$ and $A^{'}_{\mu}$, which depends on trigonometric functions. 

\section{\it Potential barriers between vacuum solutions  }

New phenomena appear when the Abelian field is switched on $\vec{H} \neq 0$ in (\ref{choansatz} ),  (\ref{spacetimefields}). The  equations (\ref{choansatz}),  (\ref{consfield}), (\ref{generasol}) and (\ref{magneticsheetsolution1}) represent an exact non-perturbative solution of the YM equation in the background chromomagnetic field \cite{Savvidy:2024sv, Savvidy:2024ppd, Savvidy:2024xbe}.    The  solution is given by the sum (\ref{choansatz})  $A^a_i = -{1\over 2} F_{ij} x_{j} n^a + {1\over g} \varepsilon^{abc} n^{b} \partial_{i}n^{c} $, $A^a_0 =0$, where $n^a$ is defined by the equation (\ref{generasol}).  The solution is parametrised by two vectors $\vec{a}$ and $\vec{b}$, and the magnetic energy density has the following form \cite{Savvidy:2024sv, Savvidy:2024ppd, Savvidy:2024xbe}:
\beqa\label{trigenergydens}
\epsilon =  {1\over 4 }G^{a}_{ij} G^{a}_{ij}  &=&   {(g  \vec{H} -   \vec{a}\times \vec{b} )^2\over 2 g^2} .
\eeqa  
This means that the  magnetic energy  density ${1\over 2}H^2$  is lowered by the vacuum polarisation and the zero magnetic energy density $\epsilon$ is realised when 
\be\label{vacuumcond}
g  \vec{H}_{vac} =  \vec{a}\times \vec{b}, ~~~~~\epsilon({g  \vec{H}_{vac} })= 0.
\ee
This takes place when three vectors  $ (\vec{H}, \vec{a}, \vec{b})$ are forming an orthogonal right-oriented frame.  {\it  At the minimum (\ref{vacuumcond}) the field strength tensor vanishes $G_{ij}=0$ and  the general solution (\ref{choansatz}),  (\ref{consfield}), (\ref{generasol}),  (\ref{magneticsheetsolution1}) reduces to a  flat vacuum connection of the following form}: 
\beqa\label{flatconnection} 
A^{a}_{i} &=&   {1\over g} \left\{
\begin{array}{cccc}   
 \Big(a b y \csc f \cos \left(\frac{b y \csc f}{f'}\right) - a f' \sin \left(\frac{b y \csc f }{f'}\right)+ \frac{a b y f''  }{f'^2}  \cos f \cos \left(\frac{b y \csc f}{f'}\right) ,  \\ 
 a b y \csc f \sin \left(\frac{b y \csc f}{f'}\right) + a f' \cos \left(\frac{b y \csc f }{f'}\right)+ \frac{a b y f'' }{f'^2}  \cos f \sin \left(\frac{b y \csc f}{f'}\right), 
  -a b y  \frac{f''  \sin f }{f'^2} \Big)  ~~~~ \\
{1\over f'}\Big(-b  \cos f  \cos \left(\frac{b y \csc f }{f' }\right),-b \cos f \sin \left(\frac{b  y \csc f }{f' }\right),   b\sin f  \Big)&\\
(0,0,0)&
\end{array} \right. 
\eeqa 
Away from the vacuum flat connection (\ref{flatconnection} ) the energy  density increases quadratically (\ref{trigenergydens}). 
Here as well one can investigate the details of the potential barriers. When the Abelian part $B_{i} n^a$  of the gauge potential  (\ref{choansatz}) is present,  then the potential barrier will take the following form:
\beqa\label{barriershapeH}
 \epsilon(g H,x,y,\alpha)&=& {a^2 b^2 \over 32 g^2} \Big( (12 - 8 \alpha + 16 \alpha^2 + 32 \alpha^3 - 8(1+4 \alpha^2){g H \over a b}~)(1- {g H \over a b}) +\\
&&+ 2  (1-4 \alpha^2) ((2{g H \over a b} -3 )^2 - 4 \alpha^2) \cos f(ax) + {(1-4 \alpha^2)^2 \over \sin^2 f(ax)} + {(1-4 \alpha^2)^2  g^2 H^2 y^2\over a^2  f^{'}_x(ax)^2} \Big). \nn
\eeqa
At $H=0$ it reduces to the previous expression  (\ref{barriershape}) $\epsilon(0,x,y,\alpha)= \epsilon(x,\alpha )$ and at  $\alpha = \pm 1/2$ we have 
\be\label{min1}
\epsilon(H,x,y, \pm 1/2) = {( g H -  a b )^2\over 2 g^2} . 
\ee 
The initial  (\ref{magneticsheetsolution1}) and final (\ref{magneticsheetsolution3}) configurations at $g H  =  a b$   are flat connections (\ref{flatconnection} ) and $A^{'}_{\mu}=0$,  so that  $G_{ij}=0$, $\epsilon =0$ and  the magnetic energy barrier between these flat connections is
\beqa\label{barriershapeH}
 \epsilon(a b,x,y,\alpha)&=&  
  (1-4 \alpha^2)^2  {a^2 b^2 \over 32 g^2}    \Big( 2  \cos f(ax) + {1\over \sin^2 f(ax)} + {    b^2   y^2\over   f^{'}_x(ax)^2} \Big). 
\eeqa
It is instructive to compare the above consideration with a  topological effect that appeared due to the presence of  gauge field configurations that have non-vanishing  Chern-Pontryagin  index.  Consider the flat connections defined in \cite{Jackiw:1976pf, Callan:1976je}:  
\be\label{flatcon} 
\vec{A}_n(\vec{x}) ={i \over g} U^{-}_n(\vec{x}) \nabla U_n(\vec{x}), ~~~~ U_1(\vec{x})= {\vec{x}^2 -\lambda^2  - 2 i \lambda \vec{\sigma} \vec{x} \over \vec{x}^2 +\lambda^2}, ~~~~~U_n = U^n_1.
\ee
The values of the gauge field  (\ref{flatcon}), although gauge equivalent to $\vec{A}(x) = 0$, are not removed from the integration over the field configurations by gauge fixing procedure because they belong to different topological classes and are separated by potential barriers of the following shape  \cite{Jackiw:1976pf, Callan:1976je, Jackiw:1979ur}:
\be\label{instbarrier}
\epsilon(r, \alpha) = {1\over 4} G^a_{ij}G^a_{ij}= {6 \lambda^4   (1- 4 \alpha^2)^2 \over g^2 (r^2 +\lambda^2)^4 }.
\ee
 In the quantum theory tunnelling will occur across this barrier and the  quantum-mechanical superposition represents the  Yang Mills $\theta$ vacuum state  \cite{Jackiw:1976pf, Callan:1976je}.  In our case the flat connection (\ref{flatconnection}) can be represented in the following form 
\be\label{flatconnection1}
\vec{A}_{g\vec{H}= \vec{a}\times \vec{b}} ~=~ -{i \over g}  S^{-}  \vec{\nabla} S
\ee
 and is characterised by the vector $g\vec{H}_{vac}= \vec{a}\times \vec{b}$   which is pointing into a particular direction in the 3d-space. One can suppose that due to the tunnelling transitions between degenerate zero energy vacua (\ref{flatconnection}), (\ref{flatconnection1}) the  quantum state  
\be
\psi(\vec{A}) = \int \CG_{  \{ \phi,\theta,\chi \} }~ \psi  (A_{g\vec{H}= \vec{a}\times \vec{b}}) ~d \mu( \phi,\theta,\chi),
\ee
where $\CG_{  \{ \phi,\theta,\chi \} }$ is the operator of space rotations,  is in a quantum-mechanical superposition of states  $\psi  (A_{g\vec{H}= \vec{a}\times \vec{b}})$ which have different orientations of the vector $g\vec{H}= \vec{a}\times \vec{b}$.  Thus  the Lorentz invariance of the vacuum state would be restored  at the quantum-mechanical level. We do not know yet whether there exist  the instanton-like solutions that would induce the tunnelling transitions between these flat configurations, but is seems probable that such solutions exist, and we will return to this problem elsewhere.

A spin system that has {\it an exponential degeneracy of the vacuum state}  was constructed in \cite{Savvidy:1993ej}.  Here the parallel planes of spins represent a vacuum configuration and the total number of such vacuum configurations is $3 \times 2^{N}$ or  $ 2^{3 N}$ if $k=0$   \cite{Savvidy:2000zq, Savvidy:1993sr, Savvidy:1994sc,Savvidy:1994tf, Pietig:1996xj, Pietig:1997va}.   In recent publications this symmetry was referred to as the {\it subsystem symmetry}, and it  has exotic fracton excitations \cite{Sherrington:1975zz, 2010PhRvB..81r4303C, Vijay:2016phm}.   

I would like to thank the members of the theoretical group at the Tor Vegata University for kind hospitality and support. I would like to thank Francesco Fucito, Massimo Bianchi,  Jos\'e Francesco Morales,  Alexander Migdal, Mikhail Shaposhnikov, Yongmin Cho, Paul Romatschke, Vincenzo Branchina and Konstantin Savvidy for stimulating discussions.

\bibliographystyle{elsarticle-num}
\bibliography{magnetic_sem}

\end{document}